\documentclass{mem}
\usepackage{natbib}\usepackage{txfonts}\usepackage{balance}
\usepackage{flushend}
\usepackage{graphicx}
\usepackage[a4paper,breaklinks,pdftex]{hyperref}
\idline{75}{282}
\begin{document}

\title{
Rotation curves of galaxies in GR}

   \subtitle{}

\author{
L.\, Ciotti}

\institute{
Dept. of  Physics and Astronomy --
University of Bologna, via Piero Gobetti 92
Bologna, Italy
\email{luca.ciotti@unibo.it}
}

\authorrunning{Ciotti}

\titlerunning{Rotation curves of galaxies in GR}

\date{Received: Day Month Year; Accepted: Day Month Year}

\abstract{It has been suggested that the observed flat rotation curves
  of disk galaxies can be a peculiar effect of General Relativity (GR)
  rather than evidence for the presence of dark matter (DM) halos in
  Newtonian gravity. In \cite{ciotti22} the problem has been
  quantitatively addressed by using the well known weak-field,
  low-velocity gravitomagnetic limit of GR, for realistic exponential
  baryonic (stellar) disks.  As expected, the resulting GR and
  Newtonian rotation curves are indistinguishable, with GR corrections
  at all radii of the order of $v^2/c^2\approx 10^{-6}$.  Here we list
  some astrophysical problems that must be faced if the existence of
  DM halos is attributed to a misinterpretation of weak field effects
  of GR.
  
}
\maketitle{}

\section{Introduction}

Following Cooperstock \& Tieu (2007, see also Balasin \& Grumiller
2008) several papers (among others, see e.g., Crosta et al. 2020;
Ludwig 2021; Astesiano \& Ruggiero 2022, and references therein),
addressed the possibility that the observed flat rotation curves of
disk galaxies at large galactocentric distances might be a GR effect
characteristic of rotating systems, with no need to invoke the
presence of DM halos, as required by Newtonian gravity.  In
\cite{ciotti22} the problem has been studied by using the well known
weak-field, low-velocity gravitomagnetic expansion of GR (see, e.g.,
Mashhoon 2008; Poisson \& Will 2014; Ruggiero 2021), by considering,
for simplicity, the case of purely stellar exponential disks, with
realistic values for the mass and scale-lengths. As expected from the
small value of the ratio $v^2/c^2\approx 10^{-6}$ in galaxies (see
point T1 below), the differences between the computed Newtonian
rotation curves and the gravitomagnetic curves are everywhere
$\approx 10^{-6}$ or less, with the conclusion that in disk galaxies
GR requires DM exactly as Newtonian gravity; this conclusion has been
recently reinforced by the studies of \cite{lasenby23},
\cite{glampedakis23}, and \cite{costa23}.

In this contribution I just list some key {\it astrophysical} problems
that should be convincingly addressed by any proposed attempt to
replace DM halos in weak-field astronomical systems by GR
effects. First, I recall below two general theoretical points:

T1) Observationally, the velocity of stars in disk galaxies is of the
order of $\approx 300\,{\rm km/s}\approx 0.001\,c$, and proportionally
less in lower mass stellar systems (some of them requiring in
Newtonian gravity even larger values of DM-to-baryon ratios than disk
galaxies, see next point A4). Classically, for these systems we expect
GR corrections to Newtonian predictions of the order of
$(v/c)^2= 10^{-6}$, while in the proposed scenario such corrections
should be strong enough to produce a flat rotation curve instead of a
Keplerian decline at large radii, a correction several times larger
than the Newtonian values!  Therefore, the claim that GR in the very
weak field, low-velocity limit of real galaxies can produce dynamical
effects {\it dominating by almost an order of magnitude} with respect
to Newtonian gravity, is fully surprising, to the point that the
consequences on the existence of DM would be of secondary
importance. If confirmed, such enormous, unexpected effect would
represent a true revolution of our understanding of GR; thus the {\it
  physical origin} of such effect should be made as transparent as
possible, with the aid of simple models and a mathematical analysis
possibly restricted to the essential.
  
T2) At present, for an astronomer reading the literature, it is
disconcerting that no consensus seems to emerge about {\it what} GR
mechanism is actually responsible for the claimed result, with a wide
spectrum of suggestions ranging from gravitomagnetism, to GR delicate
effects of boundary conditions/vacuum solutions, to retarded potential
effects due to unsteady accretion of gas on galaxies (Yahalom 2021),
to gravitomagnetic dipole effects produced by pairs of rotating black
holes (Govaerts 2023).

From the astrophysical point of view, GR solutions should convincingly
address the following points:

A1) First, a warning. Purely baryonic exponential disks (the visible
component of disk galaxies) in Newtonian gravity produce reasonably
flat rotation curves in the radial range from 1 to 3 scale-lengths: a
flat rotation curve inside this region (containing more than 80\% of
the total visible mass!)  {\it does not necessarily} require the
presence of a DM halo for its explanation (see, e.g., Kalnajs 1983;
Kent 1986). DM is only {\it required} by rotation curves in the
regions probed by the rotating HI gas, well beyond the edge of the
bright optical part of the galaxy (van Albada et al. 1985; van Albada
\& Sancisi 1986; see also Chapters 20 and 9 in Bertin 2014 and 2022,
respectively). Therefore, GR solutions producing a flat rotation curve
for the {\it stellar disk} are not alternatives to DM, but are just in
accordance with the common expectation from the Newtonian limit of GR.

A2) If DM is mimicked by GR effects related to the rotation of the
baryonic component of disk galaxies, why DM halos are found to be
required also in systems with {\it very low} rotational support, such
as Clusters of Galaxies and Elliptical Galaxies (e.g., see Cappellari
2016, Chapter 10 in Bertin 2022, and references therein)?

A3) How is the theoretical point (T1) reconciled with the fact that
for systems with gravitational strong and weak lensing (a weak-field
limit of GR), DM halos {\it are} inferred, in remarkable agreement
with Newtonian predictions based on stellar dynamics and/or
hydrostatic equilibrium of hot, X-ray emitting gaseous halos (see,
e.g., Treu 2010, Chapters 7 and 8 in Kim \& Pellegrini 2012, Chapter 6
in Meneghetti 2021)? We notice that recently it has been shown that a
GR disk model used to explore the impact of GR on the rotation curves
of disk galaxies, would also produce enormously large and unobserved
lensing deflections (Galoppo et al. 2022).
  
A4) Dwarf Spheroidal Galaxies (dSph) are low-mass and very gas poor
systems, of spheroidal shape, but they appear to have very high
DM-to-baryon ratio, even higher than disk galaxies (see, e.g., Mateo
1998; Battaglia \& Nipoti 2022).  An even more dramatic case of
low-mass systems where a significant amount of DM seems to be required
is the recently discovered class of Ultra-Faint Dwarf Galaxies (UFDs,
e.g. see Belokurov 2007, Simon 2019 and references therein): why GR
effects in these systems (with $v/c$ ratios much lower than in normal
disk galaxies) should be proportionally more important? And, strictly
related:

A5) In Globular Clusters, some of them in the same velocity
dispersion/mass range of dSph/UFDs (e.g. van den Bergh 2008), DM {\it
  is not} required\footnote{As it seems for the family of the slightly
  more massive Ultra Compact Dwarf Galaxies (see. e.g., Mahani et
  al. 2021).}: why GR should behave so differently in stellar systems
of similar mass (see also Fig.~14 in Forbes et al. 2008, Fig.~4 in
Wolf et al. 2010, and Fig.~11 in Lelli et al. 2018 for other examples
of DM-to-baryon large scatter in stellar systems of similar mass)?

A6) Cosmological simulations with DM and initial conditions derived
from the observed CMB fluctuation spectrum appear to be {\it highly}
successful in reproducing the growth of the large scale structure of
the Universe.  Attempts should be made to obtain comparable results
from GR in the absence of DM, similarly to what is currently
investigated in MOND, a very well studied alternative to DM on
galactic scales. Without entering in the debate about the viability of
MOND as an alternative to DM (not the subject of this communication),
we notice that several of the points above have been in fact addressed
quite succesfully and consistently in the MOND framework (Milgrom
1983, Bekenstein \& Milgrom 1984, see also Sanders 2014, Merritt 2020,
and references therein), along a line that should be pursued also for
the case of GR in absence of DM.

A7) As well known, DM halos in disk galaxies are required to keep the
disks stable \citep{ostriker73}. Can this stabilizing effect be
replaced by GR?

\begin{acknowledgements}
  I am grateful to an anonymous referee for a constructive report. I
  also thank Giuseppe Bertin, Alister Graham, and Silvia Pellegrini,
  for useful comments.
\end{acknowledgements}

\end{document}